\documentclass[aps,notitlepage, floatfix, prb, twocolumn]{revtex4-2}
\usepackage[colorlinks=true, allcolors=blue, bookmarks=false]{hyperref}
\usepackage{amsfonts, amsmath, amssymb}
\usepackage{graphicx}

\setcitestyle{numbers,square,sort&compress}
\begin{document}

\title{Number of zero-energy eigenstates in the PXP model}

\author{Wouter Buijsman} 

\email{buijsman@post.bgu.ac.il}

\affiliation{Department of Physics, Ben-Gurion University of the Negev, Beer-Sheva 84105, Israel}

\date{\today}

\begin{abstract}
The PXP model is paradigmatic in the field of quantum many-body scars. This model has a number of zero-energy eigenstates that is exponentially large in system size. Lower bounds on the number of zero-energy eigenstates are obtained for both open and periodic (zero and $\pi$-momentum sectors) boundary conditions. These bounds are found to be tight up to system sizes accessible by numerical exact diagonalization, and can be expected to be tight in general. In addition to previous results, separate lower bounds are obtained for the spatial inversion-symmetric and inversion-antisymmetric symmetry sectors. Furthermore, the derivations improve on previous ones as these are free of assumptions.
\end{abstract}

\maketitle

\section{Introduction}
This work focuses on the PXP model \cite{Turner18, Turner18-2}. Motivated by seminal experiments on a Rydberg atom quantum simulator \cite{Bernien17}, it has been proposed as a model for a chain of interacting two-level atoms subject to the constraint that two adjacent atoms can not be simultaneously in the excited state. The model has a number of highly non-thermal eigenstates, which are typically referred to as quantum many-body scars -- akin to the ``scarred'' eigenstates observed in certain classical billiards \cite{Heller84, Ho19}. Since the discovery of quantum many-body scars in 2018, the PXP model has been the subject of intense research (see e.g. Refs.~\cite{Serbyn21, Papic21, Moudgalya21} for related reviews).

Although the PXP model is non-integrable, a number of properties have been established analytically. Among these is an exponentially (in system size) large number of zero-energy eigenstates, so-called zero modes \cite{Turner18, Turner18-2}. Since the zero modes have a degenerate energy, any linear combination of these is a zero mode as well. It has been shown that this allows one to construct an (highly excited, as the energy spectrum is symmetric around zero) non-thermal, area-law entangled zero-energy eigenstate represented by a matrix product state with finite bond dimension \cite{Lin19}. Numerical investigations indicate that the possibility to construct an area-law entangled zero-energy eigenstate holds in more general experimentally relevant deformations of this model as well \cite{Karle21}. Besides, it has been found that a large set of exact zero-energy non-thermal eigenstates can be constructed in a systematic way from two-particle dimer states \cite{Surace21}. Interestingly, the quantum many-body scars have been found to be well approximated by quasiparticle excitations on top of zero modes \cite{Lin19}.

The exponentially large number of zero-energy eigenstates is the result of an interplay between the symmetries of the model and a parity-anticommutation relation \cite{Schecter18}. A related mechanism leading to highly degenerate zero-energy eigenstates can be observed in models of Hilbert space fragmentation \cite{Sala20, Khemani20}. Besides, the PXP model maps on certain types of other models, describing for example interacting Fibonacci anyons \cite{Lesanovsky12} or the quantum Hall effect on a thin torus \cite{Moudgalya20}. Other known mechanisms resulting in a large number of zero-energy eigenstates are based on for example supersymmetry \cite{Fendley05} or topology \cite{Ivanov01} (see also Ref.~\cite{Banerjee21} for an example in the context of quantum many-body scars). 

Refs.~\cite{Turner18, Turner18-2} established lower bounds on the number of zero-energy eigenstates as a function of the system size for open boundary conditions, periodic boundary conditions in the zero-momentum sector, and periodic boundary conditions in the $\pi$-momentum sector (see Appendix \ref{app: review} for a review of these results). Besides analytical convenience, these sectors are physically of most interest as the quantum many-body scars can be found in there. The bounds are given by Fibonacci numbers, and are numerically found to be tight (the number of zero-energy eigenstates is given by the lower bound) up to accessible system sizes.

The PXP model is invariant under spatial inversions, meaning that the lower bounds on the number of zero-energy eigenstates can be decomposed in contributions from the inversion-symmetric and inversion-antisymmetric symmetry sectors. To the knowledge of the author, this decomposition has not been discussed in the literature. Besides, the known derivations for periodic boundary conditions are based on empirical observations on the structure of sequences. The aim of this work is to re-derive the above results free of assumptions, and to establish separate lower bounds on the numbers of zero-energy eigenstates for the inversion-symmetric and inversion-antisymmetric symmetry sectors. As for the previous results, up to numerically accessible system sizes all bounds are found to be tight.

The outline of this work is as follows. Sec.~\ref{sec: PXP} reviews the PXP model, its symmetries, and the mechanism leading to zero-energy eigenstates. Sec.~\ref{sec: obc} considers open boundary conditions. Following a partially similar but slightly more involved approach, Sec.~\ref{sec: pbc-0} and \ref{sec: pbc-pi} discuss periodic boundary conditions with respectively momentum zero and momentum $\pi$. A discussion of the results and an outlook is provided in Sec.~\ref{sec: discussion}.

\section{PXP model} \label{sec: PXP}
This Section discusses the PXP model and its symmetries. The PXP model describes a chain of interacting spin-$1/2$ particles subject to the constraint that two neighboring sites can not be simultaneously in the up-state. The Hamiltonian $H$ is given by
\begin{equation}
H = \sum_{i} ( 1 - \sigma^z_i) \sigma^x_{i+1} (1 - \sigma^z_{i+2}),
\end{equation}
where $\sigma_i^{x,z}$ is a Pauli matrix acting on the $i$-th spin. Let $L$ denote the total number of spins, labeled by $i=1,2,\ldots, L$. For periodic boundary conditions, the summation runs from $i=1$ to $L$, and $\sigma_i^{x,z} \equiv \sigma_{i+L}^{x,z}$. For open boundary conditions, the summation runs from $i=1$ to $L-2$, and a term $\sigma^x_1 (1 - \sigma^z_2) + (1 - \sigma^z_{L-1}) \sigma^x_L$ is added such that the first ($i=1$) and last ($i=L$) sites can be in the up state as well when the neighboring spins are in the down state. Motivated by experimental realizations \cite{Bernien17}, spin up and down are typically referred to as respectively the ground (pictorially represented by $\circ$) and excited (represented by $\bullet$) state.

For both open and periodic boundary conditions, the Hamiltonian commutes with the (unitary) spatial inversion operator
\begin{equation}
I: i \to L-i+1.
\end{equation}
The inversion operator has eigenvalues $+1$ (symmetric eigenstates) and $-1$ (antisymmetric eigenstates). For periodic boundary conditions, the Hamiltonian additionally commutes with the (unitary) translation operator
\begin{equation}
T: = i \to i+1.
\end{equation}
The translation operator has eigenvalues $\exp(i p)$ with $p$ being referred to as the \emph{momentum} of the eigenstates. For $L$ even and odd, the momentum can take the respective values
\begin{equation}
p = \left( -1 + \frac{2n}{L} \right) \pi, \qquad p = \left( -1 + \frac{2n-1}{L} \right) \pi,
\label{eq: p}
\end{equation}
with $n = 1,2,\ldots, L$. Notice that there is a zero-momentum sector ($p = 0$) for both $L$ even and odd, and there is a $\pi$-momentum sector for $L$ even. For both open and periodic boundary conditions, the Hamiltonian finally anticommutes with the parity operator
\begin{equation}
\mathcal{C} = (-1)^L \prod_{i=1}^L \sigma^z_i,
\end{equation}
which is a consequence of the observation that the number of excitations changes from even (odd) to odd (even) under its action. The parity operator has eigenvalues $+1$ and $-1$. Notice that $[I, \mathcal{C}] = 0$ and $[T, \mathcal{C}] = 0$, indicating that the anticommutation relation holds separately for each symmetry sector.

The eigenvalue equation of the (real-valued) Hamiltonian can be written in the form
\begin{equation}
\begin{pmatrix}
0 & X \\ 
 X^T & 0
\end{pmatrix}
\begin{pmatrix} 
\psi_\text{even} \\ \psi_\text{odd}
\end{pmatrix}
= E
\begin{pmatrix} 
\psi_\text{even} \\ \psi_\text{odd}
\end{pmatrix},
\label{eq: diagonalization}
\end{equation}
where $E$ denotes an energy eigenvalue and $X$ is a (depending on system size, non-square) matrix. Here, $\psi_\text{even}$ and $\psi_\text{odd}$ give the eigenstate components for the basis states with respectively an even and odd number of excitations. One observes that if $( \psi_\text{even}, \psi_\text{odd})^T$ is an eigenstate with energy $E$, then $( \psi_\text{even}, -\psi_\text{odd})^T$ is an eigenstate with energy $-E$. As these states are orthogonal, it follows that $ | \psi_\text{even} |^2 = | \psi_\text{odd} |^2 = 1/2$. The spectrum is thus symmetric around energy zero. 

Adapting the notation of Eq.~\eqref{eq: diagonalization}, the parity operator $\mathcal{C}$ takes the form
\begin{equation}
\mathcal{C} =
\begin{pmatrix}
1 & 0 \\
0 & -1
\end{pmatrix}.
\end{equation}
For $X$ of dimension $n \times m$, there are $2 \min(n,m)$ eigenstates with $| \psi_\text{even} |^2 = | \psi_\text{odd}|^2 = 1/2$. These eigenstates have a parity expectation value $\mathcal{C} = 0$. This expectation value is known as the \emph{chiral charge}. For the respective cases $n > m$ and $n < m$, there are $|n - m|$ eigenstates of the form
\begin{equation}
\begin{pmatrix}
\psi_\text{even} \\ 0
\end{pmatrix},
\qquad
\begin{pmatrix}
0 \\ \psi_\text{odd}
\end{pmatrix},
\label{eq: eigenstates}
\end{equation}
which have respectively a chiral charge $\mathcal{C} = +1$ and $\mathcal{C} = -1$. The total chiral charge $Q$ of all eigenstates is thus equal to $n - m$. Eq.~\eqref{eq: diagonalization} shows that the eigenvalues corresponding to eigenstates with chiral charge $\pm 1$ are given by $E = 0$. The number of zero-energy eigenstates $\mathcal{Z}$ is thus lower bounded by $\mathcal{Z} \ge | Q |$. The actual number can be larger due to zero-energy eigenstates for which the eigenvalue zero is not related to the anticommutation relation $\{ H, \mathcal{C} \} = 0$. This, however, requires fine-tuning of the model in most practical settings (for an exception in the context of quantum many-body scars, see Ref.~ \cite{Surace21-2}).

\section{Open boundary conditions} \label{sec: obc}
In this Section, a lower bound on the number of zero-energy eigenstates for open boundary conditions is determined. The main difference from the approach used in Refs.~\cite{Turner18, Turner18-2} is that here different particle numbers are considered separately, which allows one to decompose the number in contributions from the inversion-symmetric and inversion-antisymmetric symmetry sectors.

The number of ways $\Omega_{L,N}$ to distribute $N$ excitations over $L$ sites with the constraint that no two consecutive sites can be in the excited state (from now on, this is assumed implicitly) is given by
\begin{equation}
\Omega_{L,N} = {L-N \choose N} + {L - N \choose N - 1}.
\label{eq: Omega}
\end{equation}
The first term accounts for the number of configurations with $N$ motifs $\bullet \circ$ and $L-2N$ sites in the ground state ($\circ$). This number does not include configurations for which the last site is in the excited state ($\bullet$). The second term accounts for the remaining number of configurations with the last site in the excited state, next consisting of $N-1$ motifs $\bullet \circ$ and $L-2N+1$ sites in the ground state. Here and throughout the remainder of this work, binomial coefficients ${n \choose k}$ are set to zero when $n < 0$, $k < 0$, or $n < k$.

The total chiral charge $Q_L$ of the eigenstates is given by the number of configurations with an even number of excitations ($\mathcal{C} = +1$) minus the number of configurations with an odd number of excitations ($\mathcal{C} = -1$),
\begin{align}
Q_L 
& = \sum_{n \ge 0} \bigg( \Omega_{L,2n} \bigg) - \sum_{n \ge 0} \bigg( \Omega_{L,2n+1} \bigg) 
\label{eq: QL-0} \\
& = \frac{1}{2} \bigg( (-1)^{\lfloor (L+1)/3 \rfloor} + (-1)^{\lfloor (L+2)/3 \rfloor} \bigg),
\label{eq: QL}
\end{align}
where $\lfloor x \rfloor$ denotes the largest integer smaller than or equal to $x$. See Appendix \ref{app: derivation} for a derivation. This result states that the number of zero-energy eigenstates $\mathcal{Z}_L \ge | Q_L |$ is lower bounded by either zero or one. This bound can be tightened by taking into account the presence of the spatial inversion symmetry. 

For a given configuration $ | c \rangle$ (e.g. $c = \circ \circ \bullet \circ \bullet$) and its spatial inverse $I | c \rangle$ ($ | \bullet \circ \bullet \circ \circ \rangle$ in the example), an (unnormalized) inversion-symmetric state can be constructed as $ | c \rangle + I | c \rangle$. Provided that $ | c \rangle \neq I | c \rangle$, an inversion-antisymmetric state can be constructed as $ | c \rangle - I | c \rangle$. The difference $\Delta_{L,N} = \Omega_{L,N}^{(+)} - \Omega_{L,N}^{(-)}$ between the contributions $\Omega_{L,N}^{(+)}$ from the inversion-symmetric and $\Omega_{L,N}^{(-)}$ from the inversion-antisymmetric sector to $\Omega_{L,N}$ is thus given by the number of $L$-site configurations with $N$ excitations that is invariant under spatial inversion. In terms of $\Omega_{L,N}$ and $\Delta_{L,N}$, the total chiral charge of the eigenstates $Q^{(+)}_{L}$ in the inversion-symmetric and $Q^{(-)}_{L}$ in the inversion-antisymmetric sector can be expressed as
\begin{equation}
\begin{split}
Q^{(\pm)}_{L} 
& = \frac{1}{2} \sum_{n \ge 0} \bigg( \Omega_{L, 2n} \pm \Delta_{L, 2n} \bigg) \\
& - \frac{1}{2} \sum_{n \ge 0} \bigg( \Omega_{L, 2n+1} \pm \Delta_{L, 2n+1} \bigg).
\end{split}
\label{eq: Qpm}
\end{equation}
In order to evaluate $Q_L^{(\pm)}$, the quantity $\Delta_{L,N}$ is determined below.

The numbers of zero-energy eigenstates $\mathcal{Z}_L^{(+)}$ and $\mathcal{Z}_L^{(-)}$ for respectively the inversion-symmetric and inversion-antisymmetric symmetry sector are lower bounded by
\begin{equation}
\mathcal{Z}_L^{(\pm)} \ge |Q_L^{(\pm)}|.
\label{eq: Zpm}
\end{equation}
A lower bound on the total number of zero-energy eigenstates $\mathcal{Z}_L \ge |Q_L^{(+)}| + |Q_L^{(-)}|$ (numbers in the inversion-symmetric and inversion-antisymmetric sectors added up) can be obtained by considering the signs of $Q_L^{(\pm)}$. If $Q_L^{(+)}$ and $Q_L^{(-)}$ are of the same sign, then $|Q_L^{(+)}| + |Q_L^{(-)}| = |Q_L^{(+)} + Q_L^{(-)}|$, which is larger than $|Q_L^{(+)} - Q_L^{(-)}|$. However, when $Q_L^{(+)}$ and $Q_L^{(-)}$ are of opposite same sign, then $|Q_L^{(+)}| + |Q_L^{(-)}| = |Q_L^{(+)} - Q_L^{(-)}|$, which is larger than $|Q_L^{(+)} + Q_L^{(-)}|$. These considerations lead to
\begin{equation}
\mathcal{Z}_L \ge \max \left( \left|Q_L^{(+)} + Q_L^{(-)} \right|, \left|Q_L^{(+)} - Q_L^{(-)} \right| \right),
\label{eq: sign}
\end{equation}
with the first or the second term being the largest when $Q_L^{(+)}$ and $Q_L^{(-)}$ are respectively of equal or opposite sign. In Ref.~\cite{Turner18-2}, assuming that $Q_L^{(\pm)}$ are of opposite sign, this inequality appears in Eq.~(A8) [reprinted in this work as Eq.~\eqref{eq: Z-app}] as $\mathcal{Z}_L \ge |K_e - K_o|$, where $K_e$ and $-K_o$ correspond to respectively the contributions from the first and second line of Eq.~\eqref{eq: Qpm} to $Q_L^{(+)} - Q_L^{(-)}$.

\subsection{Even number of sites}
First suppose that the number of sites $L = 2l$ is even. For an even number $N = 2n$ of excitations, inversion-symmetric configurations are of the form $A \circ \circ (IA)$, where $A$ is an $(l-1)$-site configuration with $n$ excitations, and $IA$ is its spatial inverse (e.g. if $A = \bullet \circ \circ$, then $IA = \circ \circ \bullet$). The number of inversion-symmetric configurations is thus equal to the number of $l$-site configurations with $n$ excitations, provided that the last site is in the ground state. Following the combinatorics as outlined below Eq.~\eqref{eq: Omega}, it follows that
\begin{equation}
\Delta_{2l, 2n} = {l - n \choose n}.
\end{equation}
It is not possible to construct inversion-symmetric configurations with an even number of sites and an odd number of excitations. Thus,
\begin{equation}
\Delta_{2l, 2n+1} = 0.
\end{equation}
The quantity $|Q_L^{(+)} + Q_L^{(-)}|$ has been evaluated in Eq.~\eqref{eq: QL}. Evaluating $|Q_L^{(+)} - Q_L^{(-)}|$ using the above expressions for $\Delta_{2l,N}$ and substituting the result in Eq.~\eqref{eq: sign} gives
\begin{align}
\mathcal{Z}_{2l}
& \ge \bigg| \sum_{n \ge 0} \bigg(\Delta_{2l, 2n} - \Delta_{2l, 2n+1} \bigg) \bigg| 
\label{eq: Z-opposite} \\
& = \sum_{n \ge 0} {l-n \choose n} 
\label{eq: binomial-0} \\
& = F_{l+1}, \label{eq: binomial}
\end{align}
where $F_n$ is the $n$-th Fibonacci number, which is recursively defined through
\begin{equation}
F_l = F_{l-1} + F_{l-2}, \qquad F_0 = 0, \qquad F_1 = 1.
\label{eq: Fibonacci}
\end{equation}
The binomial identity relating Eq.~\eqref{eq: binomial-0} to Eq.~\eqref{eq: binomial} can be found e.g. below Eq.~(1.74) of Ref.~\cite{Gould72}. The Fibonacci numbers are given non-recursively by Binet's formula (see e.g. Eq.~(1.74) in Ref.~\cite{Gould72}),
\begin{equation}
F_l = \frac{\varphi^l - (-\varphi)^{-l}}{\sqrt{5}},
\label{eq: Binet}
\end{equation}
where $ \varphi = (1 + \sqrt{5}) / 2$ is the so-called \emph{golden ratio}. From this, it follows that for $l \to \infty$, one finds $F_l \sim \varphi^l / \sqrt{5}$. Lower bounds on $\mathcal{Z}_{2l}^{(\pm)}$ follow from Eq.~\eqref{eq: Zpm}. Plots of $|Q_L^{(+)}|$ and $|Q_L^{(-)}|$, as well as a qualitative discussion are provided at the end of the discussion below of $\mathcal{Z}_L^{(\pm)}$ for $L$ odd.

\subsection{Odd number of sites}
Next suppose that $L=2l+1$ is odd. For $N=2n$ even, inversion-symmetric configurations are of the form $A \circ (IA)$, where $A$ is an $l$-site configuration with $n$ excitations. Hence, the number of inversion-invariant configurations is given by the number of ways to distribute $n$ excitations over $l$ sites [see Eq.~\eqref{eq: Omega}],
\begin{equation}
\Delta_{2l+1,2n} = {l-n \choose n} + {l - n \choose n - 1}.
\end{equation}
For $N = 2n+1$ odd, inversion-symmetric configurations are of the form $A \circ \bullet \circ (IA)$, where $A$ is an $(l-1)$-site configuration with $n$ excitations. This means that the number of inversion-invariant configurations is given by the number of ways to distribute $n$ excitations over $l-1$ sites. Thus,
\begin{equation}
\Delta_{2l+1,2n+1} = {l-n \choose n}.
\end{equation}
Analog to the procedure for an even number of sites [see Eq.~\eqref{eq: sign}], the total number of zero-energy eigenstates $\mathcal{Z}_{2l+1}$ obeys
\begin{align}
\mathcal{Z}_{2l+1}
& \ge \bigg| \sum_{n \ge 0} \bigg(\Delta_{2l+1, 2n} - \Delta_{2l+1, 2n+1} \bigg) \bigg| \\
& = \sum_{n \ge 0} {l - n \choose n - 1} \\
& = F_{l},
\end{align}
where $F_l$ is the $l$-th Fibonacci number [see Eq.~\eqref{eq: Fibonacci}]. Again, lower bounds on the contributions from the inversion-symmetric and inversion-antisymmetric sectors can be obtained through Eq.~\eqref{eq: Zpm}.

\begin{figure}
\includegraphics[scale=0.9]{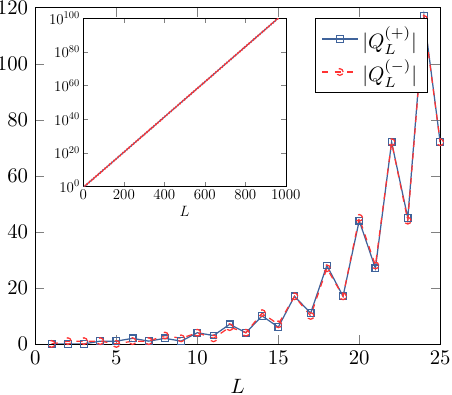}
\caption{Plots of $|Q_L^{(+)}|$ and $|Q_L^{(-)}|$ as a function of $L$ for $L=1, \dots, 25$ on a linear scale (main panel) and for $L=1, \dots, 1000$ on a logarithmic scale (inset). The curves in the main panel are close to each other, and the curves in the inset are visually indistinguishable. The data shown in the inset can be fitted by $|Q_L^{(\pm)}| \simeq \mu^{(\pm)} \varphi^{L/2}$ with $ \varphi = (1 + \sqrt{5}) / 2$ and $\mu^{(+)} = \mu^{(-)} = 0.291 \pm 0.003$ (data for $L \le 50$ excluded).}
\label{fig: obc}
\end{figure}

Fig. \ref{fig: obc} shows plots of $|Q_L^{(+)}|$ and $|Q_L^{(-)}|$ as a function of $L$. The bounds $\mathcal{Z}_L^{(\pm)} \ge |Q_L^{(\pm)}|$ have been found to be tight up to system sizes accessible by numerical exact diagonalization ($L=22$). As $|\mathcal{Z}_l^{(+)} - \mathcal{Z}_L^{(-)}| \le 1$ by Eq.~\eqref{eq: QL}, the relative difference tends to zero for large $L$. As naturally expected, $|Q_L^{(\pm)}| \simeq \mu^{(\pm)} [(1+ \sqrt{5}) /2]^{L/2}$ [see Eq.~\eqref{eq: Binet}] for large values of $L$ (the fitted prefactors $\mu^{(\pm)}$ are given in the caption).

\section{Periodic boundary conditions at zero momentum} \label{sec: pbc-0}
Zero-momentum ($p=0$) states are invariant under the action of the translation operator as the corresponding eigenvalue is given by $e^{ip} = 1$. First, the total number of orthogonal zero-momentum states (numbers for the inversion-symmetric and inversion-antisymmetric symmetry sectors added up) is determined. From a given configuration $ | c \rangle$, a translationally invariant state $ | c^{(0)} \rangle$ can be constructed as
\begin{equation}
 | c^{(0)} \rangle = \sum_{i=0}^{L-1} T^i | c \rangle.
\label{eq: c-0}
\end{equation}
Note that for some configurations (e.g. $c = \bullet \circ \circ \bullet \circ \circ$), a translationally invariant state results already from the first terms of the summation (up to $i=2$ instead of $i=5$ in the example). For a configuration with $N$ excitations and the first site in the excited state, there are at most $N-1$ unique different configurations with the first site in the excited state that can be obtained by repeatedly applying the translation operator. The number $\Phi_{L, N}$ of configurations with $N$ excitations on $L$ sites with the first site in the excited state is given by
\begin{equation}
\Phi_{L,N} = {L - N - 1\choose N - 1},
\end{equation}
see below Eq.~\eqref{eq: Omega} for a justification. For the moment ignoring `at most', the number of orthogonal zero-momentum states for given $L$ and $N$ is thus given by $\Phi_{L,N} / N$. 

As mentioned, certain configurations (e.g. $c = \bullet \circ \circ \bullet \circ \circ$) have a lower number of different configurations with a particle on the first site that can be obtained by repeatedly applying the translation operator. These configurations consist of $d$ ($d=2$ in the example) repeating motifs ($\bullet \circ \circ$ in the example). It thus follows that the number $\Phi_{L,N,d}^{(0)}$ of $L$-site configurations with $N$ excitations that consist of $d$ repeating motifs with an excitation on the first site is given by
\begin{equation}
\Phi_{L,N,d}^{(0)} =
\begin{cases}
\displaystyle {L/d - N/d - 1 \choose N/d - 1} & \text{if } (L/d),(N/d) \in \mathbb{N}, \\
0 & \text{otherwise.}
\end{cases}
\label{eq: Phi-d0}
\end{equation}
Let the number of $L$-site configurations containing $N$ excitations that consist of $d$ repeating motifs with an excitation on the first site that are not configurations of $i \times d$ ($i > 1$) repeating motifs with an excitation on the first site be given by $\tilde{\Phi}_{L,N,d}^{(0)}$. This quantity is given in recursive form by
\begin{equation}
\tilde{\Phi}_{L,N,d}^{(0)} = \Phi_{L,N,d}^{(0)} - \sum_{i > 1} \tilde{\Phi}_{L,N,i \times d}^{(0)}.
\label{eq: PhiTilde-0}
\end{equation}
From the $\tilde{\Phi}_{L,N,d}^{(0)}$ configurations, one can construct $d / N \times \tilde{\Phi}_{L,N,d}^{(0)}$ orthogonal zero-momentum states. In terms of this number, the number of orthogonal $L$-site zero-momentum states $\Omega_{L,N}^{(0)}$ with $N$ excitations is thus given by 
\begin{equation}
\Omega_{L,N}^{(0)} =
\begin{cases}
\displaystyle \frac{1}{N} \sum_{d \ge 1} d \times \tilde{\Phi}_{L,N, d}^{(0)} & \text{if } N > 0, \\
1 & \text{if } N = 0.
\end{cases}
\end{equation}
Since the basis state with zero excitations is translationally invariant, one finds $\Omega_{L,0}^{(0)} = 1$.
 
Analog to the procedure for open boundary conditions, next the difference between the number of orthogonal zero-momentum states in the inversion-symmetric and inversion-antisymmetric sectors is determined. From a zero-momentum state $ | c^{(0)} \rangle$ as given in Eq.~\eqref{eq: c-0}, a zero-momentum, inversion-symmetric state can be constructed as $ | c^{(0)} \rangle + I | c^{(0)} \rangle$. Provided that $ | c^{(0)} \rangle \neq I | c^{(0)} \rangle$, a zero-momentum, inversion-antisymmetric state can be constructed as $ | c^{(0)} \rangle - I | c^{(0)} \rangle$. The contribution to $\Omega^{(0)}_{L,N}$ from the inversion-symmetric sector is denoted by $\Omega_{L,N}^{(0, +)}$, while the contribution from the inversion-antisymmetric sector is denoted by $\Omega_{L,N}^{(0, -)}$. The difference $\Delta^{(0)}_{L,N} = \Omega_{L,N}^{(0, +)} - \Omega_{L,N}^{(0, -)}$ is given by the number of orthogonal $L$-site momentum-zero states with $N$ excitations that is invariant under spatial inversion. The total chiral charges of the eigenstates $Q_{L}^{(0,+)}$ in the inversion-symmetric and $Q_{L}^{(0,-)}$ in the inversion-anisymmetric sector are then given by
\begin{equation}
\begin{split}
Q^{(0, \pm)}_{L} 
& = \frac{1}{2} \sum_{n \ge 0} \bigg( \Omega_{L, 2n}^{(0)} \pm \Delta_{L, 2n}^{(0)} \bigg) \\
& - \frac{1}{2} \sum_{n \ge 0} \bigg( \Omega_{L, 2n+1}^{(0)} \pm \Delta_{L, 2n+1}^{(0)} \bigg),
\end{split}
\label{eq: Qpm-0}
\end{equation}
analog to Eq.~\eqref{eq: Qpm}. The corresponding numbers of zero-energy eigenstates $\mathcal{Z}^{(0,+)}_L$ in the inversion-symmetric and $\mathcal{Z}^{(0,-)}_L$ in the inversion-antisymmetric sector obey $\mathcal{Z}^{(0,\pm)}_L \ge | Q^{(0, \pm)}_L |$. Aiming to evaluate $Q^{(0, \pm)}_{L}$, the quantity $\Delta_{L.N}^{(0)}$ is evaluated below.

\subsection{Even number of sites}
First, the focus is on the case in which both $L = 2l$ and $N = 2n$ are even. Zero-momentum states as given in Eq.~\eqref{eq: c-0} that are invariant under spatial inversion are constructed out of a configuration $ | c \rangle$ for which $T^i | c \rangle = I | c \rangle$ for some $i \ge 0$. Configurations for which $ | c \rangle = I | c \rangle$ (case $i=0$) are of the form $c = \circ A \circ \circ (I A) \circ$, where $A$ is an $l-2$ site configuration with $n$ excitations, and $IA$ is its spatial inverse. Configurations for which $T | c \rangle = I | c \rangle$ (case $i=1$) are of the form $c = B \circ (IB) \circ$, where $B$ is an $l-1$ site configuration with $n$ excitations. Remark that $T I = I T^{-1}$, and that the first site of $B$ needs to be in the excited state as otherwise $B \circ$ can be written as $\circ A \circ$ by taking $A$ as $B$ with the first site removed, leading to double counting. Configurations for which $T^{2i} | c \rangle = I | c \rangle$ obey $T^i | c \rangle = I T^i | c \rangle$, and are thus covered by the case $i = 0$. Configurations for which $T^{2i+1} | c \rangle = I | c \rangle$ obey $T^i | c \rangle = I T^{i+1} | c \rangle$, and are thus covered by the case $i = 1$. Notice that, as a consequence, states constructed out of configurations that consist of repeated motifs (for which multiple values of $i$ can be found) are counted only once.

The quantity $\Delta_{2l, 2n}^{(0)}$ is given by the number of possible configurations $c = \circ A \circ \circ (I A) \circ$ and $c = B \circ (IB) \circ$ with $A$ and $B$ as defined above. This means that $\Delta_{2l, 2n}^{(0)}$ equals the number of $l$-site configurations (namely, $\circ A \circ$ or $B \circ$) with $n$ excitations, provided that the last site is in the ground state. Thus,
\begin{equation}
\Delta^{(0)}_{2l, 2n} = {l - n \choose n},
\label{eq: LeNe}
\end{equation}
see below Eq.~\eqref{eq: Omega}.

Next suppose that $N = 2n+1$ is odd. Configurations for which $T | c \rangle = I | c \rangle$ are of the form $c = A \circ \bullet \circ (IA) \circ$, where again $A$ is a configuration with $n$ excitations on $l-2$ sites. Configurations for which $T^{2i+1} | c \rangle = I | c \rangle$ are again covered by the case $i = 1$. No configurations for which $ | c \rangle = I | c \rangle$ can be found, like configurations for which $T^{2i} | c \rangle = I | c \rangle$. From this, it follows that
\begin{equation}
\Delta^{(0)}_{2l, 2n+1} = {l - n - 1 \choose n}
\label{eq: LeNo}
\end{equation}
counts the number of possible configurations ($A \circ$) on $l-1$ sites with $n$ excitations, provided that the last site is in the ground state. 

The number of zero-energy eigenstates $\mathcal{Z}^{(0)}_{2l}$ in the zero-momentum sector is lower bounded by $\mathcal{Z}^{(0)}_{2l} \ge | Q^{(0,+)}_{2l} - Q^{(0,-)}_{2l} |$, see Eq.~\eqref{eq: sign} for a justification. Analog to the procedure followed for Eq.~\eqref{eq: Z-opposite}, it follows that $\mathcal{Z}^{(0)}_{2l} \ge F_{l+1} - F_l = F_{l-1}$, where $F_l$ is the $l$-th Fibonacci number [see Eq.~\eqref{eq: Fibonacci}]. Plots of $|Q_L^{(0,+)}|$ and $|Q_L^{(0,-)}|$ are provided below the discussion of $\mathcal{Z}_L^{(0,\pm)}$ for $L$ odd.

\subsection{Odd number of sites}
Next, suppose that $L = 2l+1$ is odd. For $N=2n$ even, configurations for which $ | c \rangle = I | c \rangle$ holds are of the form $c = \circ A \circ (IA) \circ$, where $A$ is a configurations consisting of $l-1$ sites with $n$ excitations. Following the reasoning for $L$ and $N$ odd [see above Eq.~\eqref{eq: LeNo}], it follows that
\begin{equation}
\Delta^{(0)}_{2l+1, 2n} = {l - n \choose n},
\end{equation}
which gives the number of possible $l$-site configurations $\circ A$ with $n$ excitations, provided that the first site is in the ground state. For $N = 2n+1$ odd, configurations satisfying $ | c \rangle = I | c \rangle$ are of the form $c = \circ B \circ \bullet \circ (IB) \circ$ with $B$ denoting a configuration of $l-2$ sites with $n$ excitations. Following the reasoning for $L$ even and $N$ even [see above Eq.~\eqref{eq: LeNe}], it follows that
\begin{equation}
\Delta^{(0)}_{2l+1, 2n+1} = {l - n - 1 \choose n},
\end{equation}
which gives the number of possible configurations $B \circ$ on $l-1$ sites with $n$ excitations, provided that the last site is in the ground state. Analog to the case of $L$ even, it follows that $\mathcal{Z}^{(0)}_{2l+1} \ge F_{l+1} - F_l = F_{l-1}$. 

\begin{figure}
\includegraphics[scale=0.9]{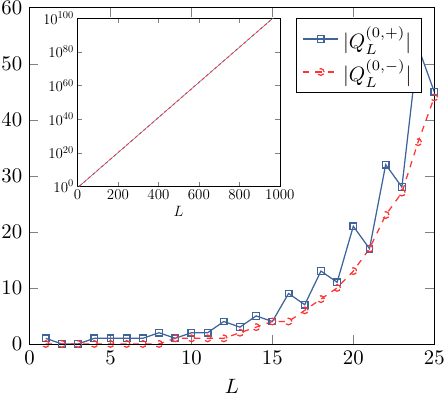}
\caption{Plots of $|Q^{(0,+)}_L|$ and $|Q^{(0,-)}_L|$ as a function of $L$ for $L=1, \dots, 25$ on a linear scale (main panel) and for $L=1, \dots, 1000$ on a logarithmic scale (inset). The curves in the inset are visually indistinguishable. The data shown in the inset can be fitted by $|Q_L^{(0,\pm)}| \simeq \mu^{(0,\pm)} \varphi^{L/2}$ with $ \varphi = (1 + \sqrt{5}) / 2$ and $\mu^{(0,+)} = 0.1272 \pm 0.0005$ and $\mu^{(0,-)} = 0.1266 \pm 0.0005$ (data for $L \le 50$ excluded).}
\label{fig: pbc-0}
\end{figure}

Fig. \ref{fig: pbc-0} shows plots of $|Q_L^{(0,+)}|$ and $|Q_L^{(0,-)}|$ as a function of $L$. The bounds $\mathcal{Z}_L^{(0, \pm)} \ge |Q_L^{(0, \pm)}|$ have been found to be tight up to numerically accessible system sizes ($L=22$). The data obeys $|Q_L^{(0, \pm)}| \simeq \mu^{(0,\pm)} [(1+ \sqrt{5}) /2]^{L/2}$ [see Eq.~\eqref{eq: Binet}] for large $L$ (the fitted prefactors $\mu^{(0,\pm)}$ are given in the caption). The bounds on the number of zero-energy eigenstates can be evaluated for single values $L \le 500$ in less than $20$ minutes of computational time on a single core of a laptop processor.

\section{Periodic boundary conditions at momentum $\pi$} \label{sec: pbc-pi}
States with momentum $\pi$ change sign under the action of the translation operator ($e^{ip} = -1$). Consequently, these states can only be found when the number of sites is even [see also Eq.~\eqref{eq: p}] and the number of excitations is at least one. As above, first the total number of orthogonal states within the sector is determined, after which it is decomposed in contributions from the inversion-symmetric and inversion-antisymmetric symmetry sectors. 

From a given configuration $ | c \rangle$ consisting of $L=2l$ sites, a $\pi$-momentum state $ | c^{(\pi)} \rangle$ can be constructed as
\begin{equation}
 | c^{(\pi)} \rangle = \sum_{i=0}^{2l-1} (-1)^i T^i | c \rangle
\end{equation}
provided that there is no $i \ge 0$ for which $ | c \rangle = T^{2i+1} | c \rangle$ as an odd number of translations induces a minus sign. From such a state, a $\pi$-momentum, inversion-symmetric state can be constructed as $ | c^{(\pi)} \rangle + I | c^{(\pi)} \rangle$ provided that $I | c^{(\pi)} \rangle \neq T | c^{(\pi)} \rangle$. A $\pi$-momentum, inversion-antisymmetric state can be constructed as $ | c^{(\pi)} \rangle - I | c^{(\pi)} \rangle$, provided that neither $I | c^{(\pi)} \rangle = | c^{(\pi)} \rangle$ nor $I | c^{(\pi)} \rangle = T | c^{(\pi)} \rangle$. If the second condition holds, an inversion-symmetric state is recovered. 

The total number $\Omega^{(\pi)}_{2l,N}$ of orthogonal $\pi$-momentum states (numbers for the inversion-symmetric and inversion-antisymmetric sectors added up) for a given number $L=2l$ of sites with $N$ excitations can be obtained by first defining $\Phi_{L,N,d}^{(\pi)}$ and $\Theta_{L,N,d}^{(\pi)}$ as the number of $L$-site configurations having $N$ excitations, consisting of $d$ (for $\Theta_{L,N,d}^{(\pi)}$, $d$ is required to be even and $L/d$ is required to be odd) repeating motifs with an excitation on the first site. Analog to Eq.~\eqref{eq: Phi-d0}, one finds
\begin{equation}
\Phi_{L,N,d}^{(\pi)} =
\begin{cases}
\displaystyle {L/d - N/d - 1 \choose N/d - 1} & \text{if } L/(2d), (N/d) \in \mathbb{N}, \\
 0 & \text{otherwise},
\end{cases}
\end{equation}
and
\begin{equation}
\Theta_{L,N,d}^{(\pi)} =
\begin{cases}
\displaystyle {L/d - N/d - 1 \choose N/d - 1} & \text{if } (L/d + 1)/2, (d/2), \vspace{-2mm} \\
& \hspace{3.5mm} (N/d) \in \mathbb{N}, \\
 0 & \text{otherwise}.
\end{cases}
\end{equation}
The respective numbers $\tilde{\Phi}_{L,N,d}^{(\pi)}$ and $\tilde{\Theta}_{L,N,d}^{(\pi)}$ of these configurations that do not consist of $i \times d$ ($i > 1$) of repeated smaller motifs with the same properties are analog to Eq.~\eqref{eq: PhiTilde-0} given by
\begin{equation}
\tilde{\Phi}_{L,N,d}^{(\pi)} = \Phi_{L,N,d}^{(\pi)} - \sum_{i > 1} \tilde{\Phi}_{L,N,i \times d}^{(\pi)}
\end{equation}
and
\begin{equation}
\tilde{\Theta}_{L,N,d}^{(\pi)} = \Theta_{L,N,d}^{(\pi)} - \sum_{i > 1} \tilde{\Theta}_{L,N,i \times d}^{(\pi)}.
\end{equation}
Taking into account that a configuration that contributes to $\tilde{\Phi}_{L,N,d}^{(\pi)}$ can consist of an even number of repeated motifs consisting of an odd number of sites (such that $|c^{(\pi)} \rangle = 0$), it follows that
\begin{equation}
\begin{split}
\Omega_{2l,N}^{(\pi)} 
& = \frac{1}{N} \sum_{d \ge 1} d \bigg[ \tilde{\Phi}_{2l,N, d}^{(\pi)} - \bigg( \Theta_{2l,N,d}^{(\pi)} - \tilde{\Theta}_{2l,N,d}^{(\pi)} \bigg) \\ 
& \times \delta \big( \Phi_{2l,N,d}^{(\pi)} > 0 \big) \bigg],
\end{split}
\end{equation}
where $\delta(\text{condition})$ equals unity if the condition is true and zero if it is not. The smallest system size for which the term involving $\Theta^{(\pi)}_{L,N,d}$ plays a role is $L=6$ due to the state constructed out of the configuration consisting of $2$ repeated motifs $\bullet \circ \circ$.

The total number $\Omega_{2l,N}^{(\pi)}$ of zero-energy eigenstates in the $\pi$-momentum sector can be decomposed in contributions $\Omega_{2l,N}^{(\pi, +)}$ from the inversion-symmetric and $\Omega_{2l,N}^{(\pi, -)}$ from the inversion-antisymmetric sector. Let $\Delta^{(\pi)}_{2l,N} = \Omega_{2l,N}^{(\pi, +)} - \Omega_{2l,N}^{(\pi, -)}$ denote the difference between these contributions. Analog to Eq.~\eqref{eq: Qpm-0}, the total chiral charges of the eigenstates $Q_{2l}^{(\pi,+)}$ in the inversion-symmetric and $Q_{2l}^{(\pi,-)}$ in the inversion-antisymmetric sector sector are given by
\begin{equation}
\begin{split}
Q^{(\pi, \pm)}_{2l} 
& = \frac{1}{2} \sum_{n \ge 1} \bigg( \Omega_{2l, 2n}^{(\pi)} \pm \Delta_{2l, 2n}^{(\pi)} \bigg) \\
& - \frac{1}{2} \sum_{n \ge 0} \bigg( \Omega_{2l, 2n+1}^{(\pi)} \pm \Delta_{2l, 2n+1}^{(\pi)} \bigg).
\end{split}
\end{equation}
The corresponding numbers of zero-energy eigenstates $\mathcal{Z}^{(\pi,\pm)}_{2l}$ are lower bounded by $\mathcal{Z}^{(\pi,\pm)}_{2l} \ge | Q^{(\pi, \pm)}_{2l} |$. The quantity $\Delta_{2l, N}^{(\pi)}$ is evaluated below.

First suppose that $N = 2n$ is even. The quantity $\Delta^{(\pi)}_{2l, 2n}$ is given by the number of states for which $I | c^{(\pi)} \rangle = | c^{(\pi)} \rangle$ minus the number of states for which $T | c^{(\pi)} \rangle = I | c^{(\pi)} \rangle$. The first contribution has been encountered before when considering the zero-momentum sector. The second contribution is due to the observation that $ | c^{(\pi)} \rangle - I | c^{\pi} \rangle$ gives an inversion-symmetric state if $T | c^{(\pi)} \rangle = I | c^{(\pi)} \rangle$. Applying the combinatorics as outlined above Eq.~\eqref{eq: LeNe} then gives
\begin{equation}
\Delta^{(\pi)}_{2l, 2n} = {l - n - 1\choose n} - {l - n \choose n}.
\end{equation}
Next suppose that $N = 2n+1$ is odd. In this case, no states for which $ | c^{(\pi)} \rangle = I | c^{(\pi)} \rangle$ can be found. The number of orthogonal states satisfying $T | c^{(\pi)} \rangle = I | c^{(\pi)} \rangle$ is the same as for the momentum-zero sector [see above Eq.~\eqref{eq: LeNo}]. One thus finds
\begin{equation}
\Delta^{(\pi)}_{2l, 2n+1} = -{l - n - 1 \choose n},
\end{equation}
where an overall minus sign is in place (see the beginning of this paragraph).

The total number of zero-energy eigenstates $\mathcal{Z}^{(\pi)}_{2l}$ in the $\pi$-momentum sector is lower bounded by $\mathcal{Z}^{(\pi)}_{2l} \ge | Q^{(\pi,+)}_{2l} - Q^{(\pi,-)}_{2l} |$, see Eq.~\eqref{eq: sign} for a justification. Analog to Eq.~\eqref{eq: Z-opposite}, it follows that $\mathcal{Z}^{(\pi)}_{2l} \ge (F_l - F_{l+1}) + F_{l} = -F_{l-1} + F_l = F_{l-2}$ for $l >0$ and $\mathcal{Z}_0^{(\pi)} = 0$. Here, again $F_l$ is the $l$-th Fibonacci number [see Eq.~\eqref{eq: Fibonacci}]. Remark that $F_{-1} = F_1 - F_0 = -1$.

\begin{figure}
\includegraphics[scale=0.9]{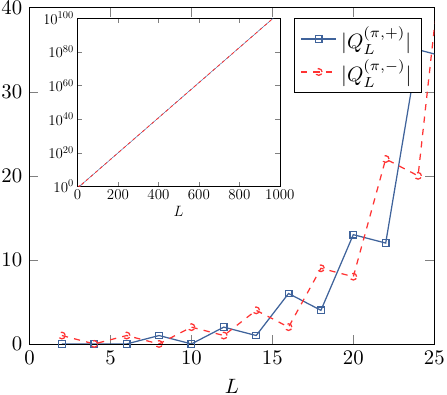}
\caption{Plots of $|Q_L^{(\pi,+)}|$ and $|Q_L^{(\pi,-)}|$ as a function of $L$ for $L=2, \dots, 25$ on a linear scale (main panel) and for $L=2, \dots, 1000$ on a logarithmic scale (inset). The curves in the inset are visually indistinguishable. The data shown in the inset can be fitted by $|Q_L^{(\pi,\pm)}| \simeq \mu^{(\pi,\pm)} \varphi^{L/2}$ with $ \varphi = (1 + \sqrt{5}) / 2$ and $\mu^{(\pi,+)} = 0.08563 \pm 0.00002$ and $\mu^{(\pi,-)} = 0.08518 \pm 0.00002$ (data for $L \le 50$ excluded).}
\label{fig: pbc-p}
\end{figure}

Fig. \ref{fig: pbc-p} shows plots of $|Q_L^{(\pi,+)}|$ and $|Q_L^{(\pi,-)}|$ as a function of $L$. The bounds $\mathcal{Z}_L^{(\pi, \pm)} \ge |Q_L^{(\pi, \pm)}|$ have been found to be tight up to numerically accessible system sizes ($L=22$). The data can be fitted by $|Q_L^{(\pi, \pm)}| \simeq \mu^{(\pi, \pm)} [(1+ \sqrt{5}) /2]^{L/2}$ [see Eq.~\eqref{eq: Binet}] for large $L$ (the fitted prefactors $\mu^{(\pi, \pm)}$ are given in the caption).

\section{Conclusions and outlook} \label{sec: discussion}
This work established lower bounds on the number of zero-energy eigenstates for the PXP model with open boundary conditions, periodic boundary conditions at zero momentum, and periodic boundary conditions at momentum $\pi$. These bounds have been decomposed in contributions from the inversion-symmetric and the inversion-antisymmetric sectors. All bounds have been found to be tight up to system sizes that can be accesses by numerical exact diagonalization, and can be expected to hold in general as exceptions would require fine-tuning of the model [see below Eq.~\eqref{eq: eigenstates}].

The results obtained in this work directly translate to deformations of the model that preserve the symmetries and the constraint that two consecutive sites can not be simultaneously in the excited state. It is natural to expect that extensions can be made to higher-dimensional generalizations of the PXP model, which are of timely interest \cite{Bluvstein21}. Of particular interest for further extensions could be the model studied in Refs.~\cite{Schecter18, Karle21}, which focus on the properties of the zero modes in the context of ergodicity breaking.

\begin{acknowledgments}
The author acknowledges support from the Kreitman School of Advanced Graduate Studies at Ben-Gurion University.
\end{acknowledgments}

\appendix
\section{Review of previous results} \label{app: review}
This Appendix reviews the results on the counting of zero-energy eigenstates obtained in Refs.~\cite{Turner18, Turner18-2}, which partially overlap with the results obtained in this work.

\subsection{Open boundary conditions}
For open boundary conditions, the starting point is 
\begin{equation}
\mathcal{Z}_L \ge | K_e - K_o |,
\label{eq: Z-app}
\end{equation}
where $\mathcal{Z}_L$ is the total number of zero-energy eigenstates (numbers for the inversion-symmetric and inversion-antisymmetric sectors added up) for systems consisting of $L$ sites, and $K_e$ ($K_o$) is the number of configurations with an even (odd) number of excitations that is inversion-symmetric [see above Eq.~\eqref{eq: sign}]. For $L=4$, as an example, the configurations $\bullet \circ \circ \bullet$ and $\circ \circ \circ \circ$ contribute to $K_e$, while $K_o = 0$.

First consider $L = 2l$ even. For any configuration $A$ of length $l-1$, there is a corresponding inversion-symmetric configuration of length $2l$ given by $A \circ \circ (IA)$. As in the main text, here $IA$ denotes is the inverse of $A$ [see above Eq.~\eqref{eq: LeNo}]. The number of possible configurations $A$ is given by $F_{l+1}$, where $F_l$ denotes the $l$-th Fibonacci number [see Eq.~{\eqref{eq: Fibonacci}]. Noting that every element of $K_e$ is of the form $A \circ \circ (IA)$ and $K_0 = 0$, it follows that $\mathcal{Z}_{2l} \ge F_{l+1}$. Next consider $L = 2l+1$ odd. Elements of $K_e$ are of the form $A \circ (IA)$, where $A$ is a configuration with $l$ sites. Elements of $K_o$ are of the form $B \circ \bullet \circ (IB)$, where $B$ is a configuration of length $l-1$. Following the same reasoning, one finds $K_e = F_{l+2}$ and $K_o = F_{l+1}$, leading to $\mathcal{Z}_{2l+1} \ge F_l$.

\subsection{Periodic boundary conditions}
The derivation for periodic boundary conditions focuses on the zero momentum sector. First consider $L = 2l+1$ odd. For an odd number of excitations, translationally invariant states [see Eq.~\eqref{eq: c-0}] are constructed from a basis state $ | c \rangle$ of the form $c = \circ A \circ \bullet \circ (IA) \circ$, where $A$ is a configuration consisting of $l-2$ sites. It is shown that the number of orthogonal translationally invariant, inversion symmetric states is given by the number of possible configurations $A$. It thus follows that $K_o = F_l$. For an even number of excitations, inversion-symmetric, translationally invariant states are constructed from a basis state $ | c \rangle$ of the form $c = \circ B \circ (IB) \circ$, where $B$ is a configuration consisting of $l-1$ sites. By the same reasoning, it follows that $K_e = F_{l+1}$. Substituting these results in Eq.~\eqref{eq: Z-app} gives $\mathcal{Z}_{2l+1} \ge F_{l-1}$. Note that these results are consistent with what is found in this work, although here no decomposition in contributions from the inversion-symmetric and inversion-antisymmetric sectors can be extracted.

Next consider $L = 2l$ even. For $N$ odd, inversion-symmetric zero-momentum states are constructed from a configuration $ | c \rangle$ of the form $c = \circ A \circ \bullet \circ (IA) \circ$, where $A$ is of length $l-2$. By the same reasoning as above, it follows that $K_o = F_l$. For $N$ even, the number of inversion-symmetric zero momentum states can not be obtained by the reasoning used before [also see above Eq.~\eqref{eq: LeNe}]. Let $K = K_e + K_o$, such that Eq.~\eqref{eq: Z-app} can be rewritten as
\begin{equation}
\mathcal{Z}_L \ge | K - 2 K_o |.
\label{eq: ZK}
\end{equation}
Let $M_e$ and $M_o$ denote the numbers of inversion-symmetric and inversion-antisymmetric states composed out of two configurations with respectively an even and odd number of excitations, and let $M = M_e + M_o$. As an example, for $L=4$ the (unnormalized) states $ | \bullet \circ \bullet \circ \rangle \pm | \circ \bullet \circ \bullet \rangle$ contribute to $M_e$ (among others), while the states $ | \bullet \circ \circ \circ \rangle \pm | \circ \circ \circ \bullet \rangle$ contribute to $M_o$. Trivially, $K = 2(M+K) - (2M+K)$. As a function of $L$, it is empirically observed that the sequences $2M+K$ and $M+K$ have respective generating functions $f(x)$ and $g(x)$ given by
\begin{equation}
f(x) = \sum_{k \ge 1} \frac{ \phi(k)}{k} \ln \frac{1}{1 - x^k (1+x^k)}
\end{equation}
and
\begin{equation}
\begin{split}
g(x) 
& = \frac{1}{2} \sum_{k \ge 1} \bigg( \frac{ \phi(k)}{k} \ln \frac{1}{1 - x^k (1 + x^k)} \bigg) \\
& - \frac{1}{2} \frac{(1+x)(1+x^2)}{x^4 + x^2 - 1},
\end{split}
\end{equation}
where $ \phi(k)$ is the Euler totient function giving the number of positive integers up to $k$ that are relatively prime to $k$. The sequence $2(M+K) - (2M+K)$ as a function of $L$ thus has a generating function given by $2g(x) - f(x)$. By recognizing the generating function of the Fibonacci sequence, it follows that $2(M+K) - (2M+K) = F_{\lfloor N/2 \rfloor + 2}$. Hence, $K_e = F_{l+1}$ and thus $\mathcal{Z}_{2l+1} \ge F_{l-1}$ by Eq.~\eqref{eq: ZK}. As for open boundary conditions, this result is consistent with what is found in this work. Note that also here no decomposition in contributions from the inversion-symmetric and inversion-antisymmetric sectors can be extracted.

For the momentum $\pi$ sector, it is mentioned that the number $\mathcal{Z}^{(\pi)}_{2l}$ of zero-energy eigenstates is lower bounded by $\mathcal{Z}^{(\pi)}_{2l} \ge F_{l-2}$, also consistent with what is found in this work.

\section{Derivation of Eq.~\eqref{eq: QL}} \label{app: derivation}
This Appendix provides a derivation of Eq.~\eqref{eq: QL}. From the standard binomial identity
\begin{equation}
{n \choose k} = {n-1 \choose k} + {n-1 \choose k-1}
\end{equation}
it follows by substituting $n = (L+1) - N$ and $k = N$ that
\begin{equation}
{L-N \choose N-1} = {L-N+1 \choose N} - {L-N \choose N}.
\end{equation}
Substituting this result in Eq.~\eqref{eq: QL-0}, by a slight change of notation one obtains
\begin{equation}
Q_L = \sum_{N \ge 0} (-1)^N {L-N+1 \choose N}.
\end{equation}
This quantity can be evaluated through the binomial identity (1.75) of Ref.~\cite{Gould72},
\begin{equation}
\sum_{k=0}^{\lfloor n/2 \rfloor}(-1)^k {n-k \choose k} = \frac{1}{2} \bigg( (-1)^{\lfloor n/3 \rfloor} + (-1)^{\lfloor (n+1)/3 \rfloor} \bigg).
\end{equation}
Substituting $n = L+1$ and $k = N$ directly gives the desired result.

\nocite{apsrev42Control}
\bibliography{references}

\end{document}